\documentclass[%
 reprint,
 amsmath,amssymb,
 aps,
]{revtex4-1}

\usepackage{graphicx}
\usepackage{dcolumn}
\usepackage{bm}
\usepackage{textgreek}

\begin{document}

\preprint{APS/123-QED}

\title{
Active surface flows accelerate the coarsening of lipid membrane domains
}

\author{Daniel P. Arnold}
\email{equal contribution}
\author{Aakanksha Gubbala$^*$}

 \author{Sho C. Takatori}
 \email{stakatori@ucsb.edu}
\affiliation{
 Department of Chemical Engineering, University of California, Santa Barbara, Santa Barbara, CA 93106
}

\begin{abstract}

Phase separation of multicomponent lipid membranes is characterized by the nucleation and coarsening of circular membrane domains that grow slowly in time as $\sim t^{1/3}$, following classical theories of coalescence and Ostwald ripening.
In this work, we study the coarsening kinetics of phase-separating lipid membranes subjected to nonequilibrium forces and flows transmitted by motor-driven gliding actin filaments. 
We experimentally observe that the activity-induced surface flows trigger rapid coarsening of non-circular membrane domains that grow as $\sim t^{2/3}$, a 2$\times$ acceleration in the growth exponent compared to passive coalescence and Ostwald ripening. 
We analyze these results by developing analytical theories based on the Smoluchowski coagulation model and the phase field model to predict the domain growth in the presence of active flows. 
Our work demonstrates that active matter forces may be used to control the growth and morphology of membrane domains driven out of equilibrium. 

\end{abstract}

\maketitle

The dynamic interplay between nonequilibrium forces and membrane surfaces plays an essential role in many physical processes in living systems. 
For example, molecular motors and the cytoskeleton inside living cells generate active forces on the cell membrane, allowing cells to bend, flow, and stretch the cell surface \cite{Alberts2007,Phillips2012}. 
Reconstituted multicomponent lipid membranes can also phase separate into macroscopic domains along the membrane \cite{Veatch2003}. 
While the nucleation and coarsening kinetics of lipid membrane domains at thermodynamic equilibrium is well established \cite{Lifshitz1961, Camley2011, Camley2010, Stanich2013, Frolov2006}, we have little understanding of how membrane domains grow when subjected to nonequilibrium forces and flows, such as those that might be generated by the actin cytoskeleton.

\begin{figure}
    \centering
    \includegraphics[width=\linewidth]{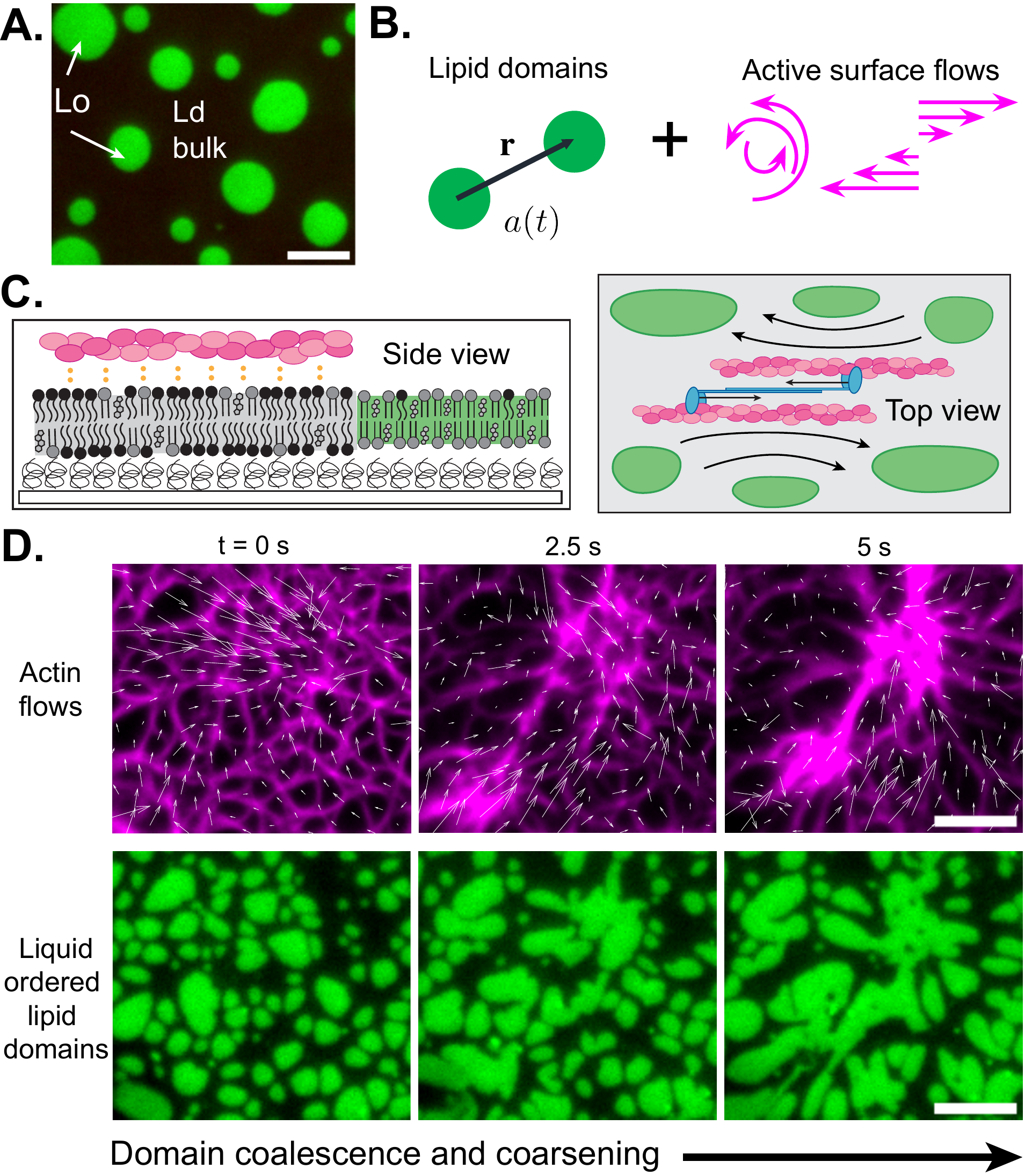}
    \caption{
    (A)
    Planar lipid bilayers phase-separate into liquid-ordered (Lo) and liquid-disordered (Ld) phases.
    (B)
    In this work, we study the growth of lipid domains of size $a$ under 2D flows generated by internal ``active'' forcing.
    (C)
    (\emph{Left}) In our experiments, actin is bound to the Ld phase, avoiding the Lo phase.
    (\emph{Right}) Actomyosin contraction internally drives lipid membrane flows.
    (D) Time-lapse images of actin (magenta) contraction with vectors calculated from particle image velocimetry overlaid (\emph{top}), and Lo lipid domains growing in time (green, \emph{bottom}).
    Scale bar is 5 \textmu m.
    }
    \label{fig:diagram}
\end{figure}

In this Letter, we experimentally and theoretically study the effect of internally-driven 2D active flows on the rate of domain coarsening in phase-separated lipid bilayers.
In the absence of active flows (Fig.~\ref{fig:diagram}A, Supplemental Movie S1), the lipids form circular, mesoscopic domains that grow slowly with time, consistent with prior work on giant vesicles \cite{Stanich2013}.
We find that actin and myosin create internally-driven surface flows, which couple with domains (Fig.~\ref{fig:diagram}B) and drive coarsening and growth according to much faster dynamics than passive systems.
In passive systems, the scaling exponent of $\alpha=1/3$ is well-established for domains of size $a$ that grow with time $t$ according to $a\sim t^\alpha$, for both coalescence and Ostwald ripening mechanisms \cite{Stanich2013,Camley2011,Wagner1961,Lifshitz1961,Siggia1979}.

Previously, it has been shown that linear, externally-imposed flows can deform domains and accelerate coarsening \cite{ohta_computer_1990,shou_ordering_2000,mcleish_coarsening_2003}. Here we describe a system in which surface-adsorbed active matter, in the form of an actomyosin cortex, internally drives lipid flows, which lead to rapid domain growth. Using Cahn-Hilliard simulations, we show that the in-plane flows created by actin increase $\alpha$ by more than a factor of 2 at moderate P\'eclet numbers of $10^{-3}$ to $10^{-2}$. We compare these results to analytical theory and simulations describing domain growth under simple shear flow, finding that the behavior of flow-based mechanisms is consistent with the experimental results for actin-driven coarsening. 

\emph{Experiments ---}
A planar lipid bilayer is deposited on a cushioned glass coverslip treated with polymer and proteins, to prevent kinetic arrest of domains due to friction with rough substrates (see Supplemental Information for detailed methods).
Briefly, silica coverslips are treated with chlorotrimethylsilane before heavy meromyosin and polylysine-grafted polyethylene glycol (PLL-g-PEG) are adsorbed to the surface.
Giant unilamellar vesicles (GUVs) are formed at 50\textdegree C with 45\% dioleoylphosphatidylcholine (DOPC), 35\% dipalmitoylphosphatidylcholine (DPPC), 15\% cholesterol, and 5\% dioleoyl-3-trimethylammonium propane (DOTAP) using established electroformation methods \cite{Angelova1986}, and then allowed to rupture on the treated coverslip, creating a planar bilayer.
At room temperature, this lipid composition phase-separates into a continuous liquid-disordered (Ld) phase containing dispersed liquid-ordered (Lo) domains (Fig.~\ref{fig:diagram}A) \cite{Veatch2003}.
Small amounts of lipid dyes are added to each phase for fluorescence imaging.

Bilayers are heated to 37\textdegree C and decorated with filamentous actin (F-actin), a negatively-charged cytoskeletal protein, which adsorbs to the bilayer via electrostatic attraction to positively-charged DOTAP enriched in the Ld phase (Fig.~\ref{fig:diagram}C, \emph{left}) \cite{Schroer2020}.
Myosin II motor proteins are added in a rigor (ATP-free) state, causing them to crosslink actin but not apply any contractile force (Fig.~\ref{fig:diagram}C, \emph{right}).
Upon quenching the system to room temperature, the Lo domains re-form and actin is sequestered into the Ld phase (Fig.~\ref{fig:diagram}C, D).
Unlike the circular domains in Fig.~\ref{fig:diagram}A, the domains in Fig.~\ref{fig:diagram}D initially adopt a morphology characterized by sharp corners and elongated edges, as they have low line tension and are forced to conform to actin bundles \cite{Tian2007, Baumgart2003}.
These actin-constrained domains reach a steady size and do not appreciably grow over tens of minutes.

Upon introducing ATP, we observe rapid actomyosin contraction, along with simultaneous elongation and growth of lipid domains (Fig.~\ref{fig:diagram}D, Supplemental Movie S2).
The most dramatic contraction is often very short-lived ($<$ 5 s) as the actin forms contractile clusters and generates surface flows that rapidly coarsen the Lo domains.
Fig.~\ref{fig:diagram}D (\emph{top}) shows the actin flow field generated by particle image velocimetry (PIV), and confirms that in-plane actin flows are directed toward an apparent sink in the upper right quadrant of the image.

\begin{figure}
    \centering
    \includegraphics{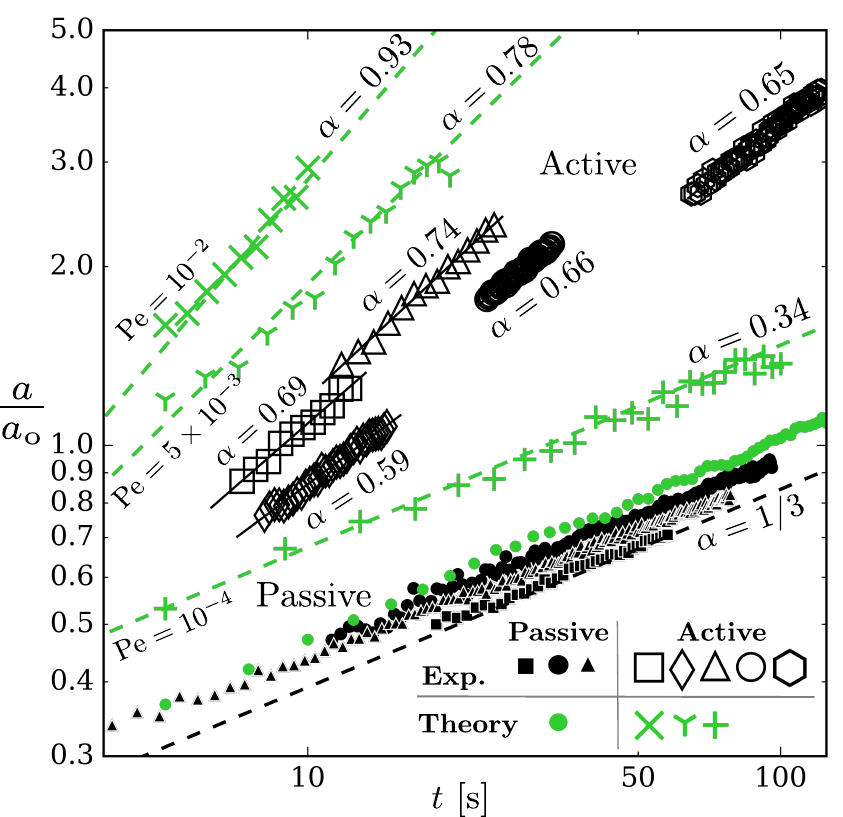}
    \caption{
    Liquid-ordered (Lo) lipid domain size, $a(t)$, is plotted as a function of time, $t$, for experiments (black) and Cahn-Hilliard numerical calculations (green), with growth exponents $\alpha$ presented for each: $a \sim t^\alpha$.
    Closed symbols represent passive domain growth in the absence of flow, with circles and triangles representing three independent passive experiments.
    Passive domains grow as $\alpha=1/3$ (dashed black line), consistent with classical theories on coalescence and Ostwald ripening.
    Open symbols represent active domain growth in the presence of 2D flow, either imposed experimentally via actomyosin contraction, or incorporated theoretically using particle image velocimetry from experiments (Fig.~\ref{fig:diagram}D).
    Five independent active experiments are presented, along with simulations using three different P\'eclet numbers (Pe).
    Each data set is rescaled by a different factor $a_0$ to capture the power law scaling exponents.
    }
    \label{fig:domain_growth}
\end{figure}

We threshold time-lapse images of the domains and track the area/perimeter ratio of the resulting binary images as a metric for characteristic domain size $a(t)$, consistent with prior experimental work on lipid domain growth \cite{Stanich2013}.
This definition of the characteristic domain size is consistent with the alternative definition based on the first moment of the static structure factor, $a(t) \sim \left( \int k S(k, t) \ \mathrm{d}k / \int S(k, t) \mathrm{d}k \right)^{-1}$, where $k$ is the wave vector (see Supplemental Information).
We fit the resulting data to the equation $a(t)=(At+B)^\alpha$ before re-scaling the time to give the form $a\sim t^\alpha$ (see Supplemental Methods).
Figure \ref{fig:domain_growth} shows the growth of $a(t)$ over time for five independent experiments with actomyosin activity (black open symbols). 
We find growth exponents range from $\alpha =$ 0.59 to 0.74 for these domains under internally-driven, active surface flows.

We measure the active coarsening rate across five different membrane compositions and actin densities, without observing a strong correlation between coarsening rate and composition (Supplemental Fig. S1) or actin density (Supplemental Fig. S2) over the range tested.
We perform an additional control experiment in which we quench the bilayer and add ATP simultaneously; we observe the passive scaling of $\alpha=1/3$ at early times when the Lo domains are small compared to the actin mesh size ($\approx$1 \textmu m), while at later times, larger domains are driven to grow more rapidly (see Supplemental Fig. S3).

We compare these active data to control experiments in which lipid bilayers without actin are heated above the miscibility temperature, and then allowed to cool so that the domains can re-form and coarsen (Supplemental Movie S1).
Figure \ref{fig:domain_growth} (black closed symbols) shows that passive membranes recover the scaling of $\alpha=1/3$, which applies to both coalescence and evaporation/condensation mechanisms of coarsening and has been well-established for membrane domains in both experimental and theoretical work \cite{Stanich2013,Camley2011,Wagner1961,Lifshitz1961} (see Supplemental Information for derivation).
We perform these passive experiments with five different lipid compositions, including critical mixtures, obtaining scaling results consistent with those of Stanich et al. on GUVs (Supplemental Figs. S1 and S4) \cite{Stanich2013}.
We hypothesize that actin contraction generates in-plane forces in the membrane, which drive the lipid domains to grow at an accelerated rate.

\emph{Theory --- }
To evaluate our hypothesis that active convection can accelerate domain coarsening, we use a Cahn-Hilliard model to evolve a phase-separating 2D system under surface flows.
We evolve a concentration order parameter, $\phi(\mathbf{x}, t)$, using the Cahn-Hilliard equation \cite{cahn_phase_1965}, which is commonly used to study coarsening in binary mixtures \cite{inguva_continuum-scale_2021, watson_coarsening_2003,o_naraigh_bubbles_2007,konig_two-dimensional_2021}. 
The Cahn-Hilliard model predicts the passive growth exponent $\alpha=1/3$ (Fig.~\ref{fig:domain_growth}, green closed circles and Supplemental Movie S3), which is consistent with prior work on Ostwald ripening in the absence of flow \cite{bray_lifshitz-slyozov_1995} (see Supplemental Information). 

To analyze the effect of 2D active flows on domain coarsening, we present the following non-dimensional Cahn-Hilliard equation in Fourier space:
\begin{equation}\label{eq:fourier_cahn_hilliard}
    \frac{\mathrm{d}\phi_k}{\mathrm{d}t} + \mathrm{Pe}\,\mathcal{F}\{\mathbf{v}\cdot\nabla\phi\} = -k^2 \mathcal{F}\bigg\{\frac{\delta f}{\delta\phi}\bigg\} - k^4 \phi_k,
\end{equation}
where $\phi_k$ is the Fourier transform of $\phi(\mathbf{x}, t)$, $k$ is the wave vector, $\mathbf{v}$ is the nondimensional convective surface velocity generated by the actin filaments, the P\'eclet number $\mathrm{Pe} \equiv \dot{\gamma}\kappa/M$, $\dot{\gamma}$ is the rate of strain, $M$ is lipid mobility, and $\kappa$ is a surface tension parameter.
We model the bulk free energy, $f\left[\phi(\mathbf{x}, t)\right] = \phi^4/4-\phi^2/2$, a double-well potential where the concentration of pure phases is $\phi=\pm 1$.
We add a convective term $\mathbf{v} \cdot \nabla \phi$ to impose surface flows on the phase-separating system, testing the effect of our hypothesized actin-induced surface convection on domain coarsening.
We use this model to obtain a mechanistic understanding of how active convection impacts the kinetics of domain growth.
Hydrodynamic effects of the fluid and the bulk Ld phase may also be included, but we omit them here because surface tension (or line tension) driven flows that cause accelerated coarsening mechanisms are only significant for near-critical point quenches where the minority phase is elongated and/or interconnected \cite{Siggia1979, Tree2017}.

We numerically solve the Cahn-Hilliard model using pseudo-spectral methods with periodic boundary conditions, starting from an initial state, $\phi_0$, with uniform noise.
To avoid any boundary artifacts, we restricted our analysis to the interior of the simulation box (see Supplemental Fig. S6).
To corroborate our experimental results, we impose the surface flow field obtained by PIV analysis of actin (Fig.~\ref{fig:diagram}D, white arrows).
We fix Cahn-Hilliard parameters $M$ and $\kappa$ based on the pixel resolution of the camera, while varying Pe to evaluate the effect of active flows on growth rate (Supplemental Movie S4).
Fig. \ref{fig:domain_growth} (green symbols) shows that domains coarsen rapidly for the strongest flows ($\alpha=0.93$ for $\mathrm{Pe}=10^{-2}$) while weaker flows effectively act as noise and recover passive scaling ($\alpha=0.34$ for $\mathrm{Pe}=10^{-4}$).
The experimentally-derived scaling exponents of $\alpha=0.59-0.74$ lie between those for $\mathrm{Pe}=10^{-4}$ and $\mathrm{Pe}=5\times 10^{-3}$ in the numerical solutions.
These results are consistent with an estimate of the molecular P\'eclet number based on literature values, $\approx 10^{-4}-10^{-3}$, using $\approx 1$ nm lipids with diffusivity $1-10$ \textmu m\textsuperscript{2}/s under surface flows with velocity $\approx 1$ \textmu m/s \cite{Seu2007,Machan2010}.

In addition to the Ostwald ripening mechanism in the Cahn-Hilliard model, we note that a different model of phase coarsening based on the Smoluchowski coagulation model predicts a $\alpha = 2/3$ growth for domains subjected to weak shear flows.
The Smoluchowski coagulation model is commonly used to predict Brownian flocculation of colloids \cite{Russel1989} and diffusion-reaction dynamics of macromolecules \cite{Fredrickson1996}, and is conceptually distinct from the molecular mechanisms driving coarsening in the Cahn-Hilliard equation. 
In the Smoluchowski perspective, the lipid domains are modeled as macroscopic colloids of fixed size that merge upon contact (i.e., dimerize via an infinitely fast chemical reaction upon contact) 
\cite{Siggia1979}.
Thus, although they both predict $\alpha = 1/3$ passive scaling, the Smoluchowski model assumes domain growth via coalescence as opposed to Ostwald ripening \cite{camley_dynamic_2011,Frolov2006}.

To obtain the enhanced scaling in shear flow under the Smoluchowski perspective, we consider the conservation of the number density of singlet domains:
\begin{equation}
\frac{\mathrm{d}n}{\mathrm{d}t}+J=0,
\label{eq:smoluchowskiIVP}
\end{equation}
where $J$ is a sink that captures the merging of singlet domains to dimerized domains,
\begin{equation} 
    J = -\frac{D_\mathrm{c}}{a} n^2 \oint \displaylimits_{r=2a} \mathbf{n} \cdot \left( \mathrm{Pe_c} \mathbf{v} -\nabla_r \ln g +\nabla_r V \right) g \ \mathrm{d}\ell. \label{eq:Smol_B}
\end{equation}
The sink depends on the pair distribution of domains, $g(\mathbf{r};\mathrm{Pe_c})$, $\mathbf{v}$ is a nondimensional velocity field, $\mathbf{n}$ is a unit vector normal to the domain surface, $V$ is the nondimensional pair potential, $\nabla_r$ is the gradient operator relative to the center of a domain, and $\mathrm{Pe_c} \equiv \dot{\gamma} a^2 / D_\mathrm{c}$ is the P\'eclet number based upon treating the domains as colloids ($D_\mathrm{c}$ is the domain diffusivity).
Assuming that the pair distribution quickly reaches a steady state at all times compared to the decay of the singlet density (quasi-static approximation), we solve for $g(\mathbf{r};\mathrm{Pe_c})$ with boundary conditions $g=0$ at contact and $g=1$ far away.
The $g=0$ condition at contact assumes an instantaneously fast merging of two singlet domains, although a finite reaction time is also straightforward to implement.
Hydrodynamic interactions between the domains may also be included in the pair distribution problem, but we omit them here because lubrication flows can generally only slow down the collision rates between membrane domains \cite{Oppenheimer2017,Siggia1979}.

Eqs. \ref{eq:smoluchowskiIVP} and \ref{eq:Smol_B} show that the singlet density decreases by a second-order reaction, $J = k^{\mathrm{eff}} n^2$, where $k^{\mathrm{eff}}$ is interpreted as the transport-limited effective reaction rate constant.
Assuming the domains are circular with no bulk interaction ($V = 0$) subjected to a simple shear flow, a perturbation analysis at small $\mathrm{Pe_c}$ gives
\begin{equation} \label{eq:Smol_k}
    k^\mathrm{eff} = \frac{\pi k_\mathrm{B}T}{8\eta a \ln\left(\sqrt{\frac{N\pi}{4\phi_\mathrm{A}}}\right)} \left( 1 + C \mathrm{Pe_c}^{1/2} \right)
\end{equation}
where $\eta$ is the bulk solvent viscosity, $N$ is the number of singlet domains at time $t=0$, $\phi_\mathrm{A}$ is the domain area fraction, and $C$ is a numerical prefactor (see Supplemental Information).
We have used the Saffman-Delbr\"{u}ck diffusivity for large domains compared to the Saffman-Delbr\"{u}ck length \cite{Saffman1975}; the membrane is embedded within a bulk 3D fluid, and the quasi-2D geometry is critical for obtaining the correct scaling \cite{Ramachandran2010, Stanich2013,Frolov2006}.
For a constant domain area fraction $\phi_A = n \pi a^2$, Eqs.~\ref{eq:smoluchowskiIVP}-\ref{eq:Smol_k} yield a domain growth scaling of $a(t) \sim t^{1/3}$ in the absence of flow ($\mathrm{Pe_c} = 0$), and an enhancement in the presence of weak shear flows, $a(t) \sim t^{2/3}$.
Therefore, both the coalescence-based Smoluchowski model and the ripening-based Cahn-Hilliard model predict a similar enhancement to the growth of domains with active convection; both mechanisms are present in the experiments.

\begin{figure}
    \centering
    \includegraphics{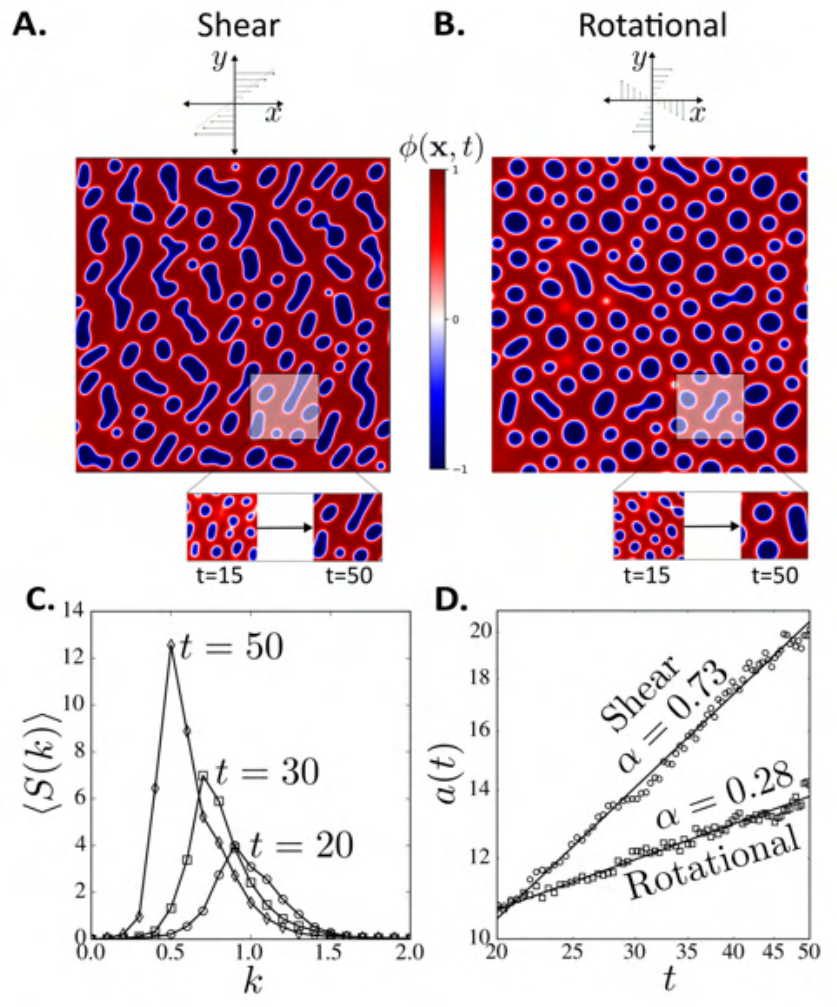}
    \caption{
    General linear flows modulate the growth and morphology of phase-separating domains. 
    Numerical solution snapshots of phase separation for Eq.~\ref{eq:fourier_cahn_hilliard} in (A) shear flow and (B) rotational flow. Inset snapshots show the evolution of the domains. 
    (C) Static structure factor for domains under shear flow.
    (D) Shear flow increases the domain growth rate, whereas rotational flow maintains the same scaling as passive Ostwald ripening, $\approx t^{1/3}$. The parameters shown here correspond to ($M$, $\kappa$, $\dot{\gamma}$)=(1, 0.25, 0.04), or $\mathrm{Pe}=10^{-2}$.
    }
    \label{fig:linear_flows}
\end{figure}

Returning to the Cahn-Hilliard model (Eq.~\ref{eq:fourier_cahn_hilliard}), we also considered simple toy models of the surface velocity to gain a more mechanistic understanding of the effect of flows on domain growth. 
We consider a general 2D linear flow $\mathbf{v}(\mathbf{x})=\mathbf{G}\cdot \mathbf{x}$ where $\mathbf{G}$ is a gradient tensor.
We compute Eq.~\ref{eq:fourier_cahn_hilliard} for shear and rotational flow fields and measure the domain size and structure as a function of time. In shear flow, we observed frequent domain merging and elongation along the extensional axis of shear (Fig.~\ref{fig:linear_flows}A, Supplemental Movie S5).
Fig. \ref{fig:linear_flows}C shows that as the static structure factor evolves in time, the magnitudes of the peaks grow as the peaks shift toward lower wave vectors.

Conversely, in rotational flow, the domains remain approximately circular, similar to the passive case (Fig.~\ref{fig:linear_flows}B, Supplemental Movie S5).
In Fig.~\ref{fig:linear_flows}D, we compare the rate of domain growth for the two linear flows, finding that at high Pe $=10^{-2}$, shear flow accelerates domain growth ($\alpha=0.73$) while rotational flow essentially recovers passive scaling ($\alpha=0.28$).
This result is consistent with the physical intuition that rigid body rotation about the center of the system should not alter the frequency of collisions between the domains.
Our analysis provides further physical insights into our experiments as any arbitrary linear flow may be constructed from linear combinations of shear flow and rotational flow.
For example, extensional flow is a sum of shear flow and rotational flow, has features that are similar to the flows observed in our experiments, and also yields $\alpha \approx 0.73$ scaling.

Both the Smoluchowski and Cahn-Hilliard models predict a $\alpha \approx 2/3$ enhancement in simple shear flow, but no enhancement in rotational flow.
We note that coalescence is an important mechanism of growth in the experiments due to the large colloidal P\'eclet number $\mathrm{Pe_c} \equiv \dot{\gamma} a^2 / D_\mathrm{c} \approx 10$, compared to the molecular P\'eclet considered in the Cahn-Hilliard model,  $\mathrm{Pe} \equiv \dot{\gamma}\kappa/M \approx 10^{-4} - 10^{-2}$.

While numerical solutions validate that 2D flows accelerate domain growth, we can obtain further mechanistic insight by linearizing the Cahn-Hilliard equation and obtaining an analytical approximation of $\alpha$ at early times.
For the case of simple shear flow, perturbation analysis at small Pe yields $a(t) \sim  t^{1/4} + \mathrm{Pe}\, t^{5/4} + \mathcal{O}(\mathrm{Pe}^2)$ (see Supplemental Information). 
Note that the passive scaling obtained from the linearized equation is $t^{1/4}$, which is different from the $t^{1/3}$ scaling obtained from the numerical solution to the fully nonlinear Eq.~\ref{eq:fourier_cahn_hilliard}.
This is consistent with prior studies of the Cahn-Hilliard equation in the context of the kinetics of Ostwald ripening \cite{konig_two-dimensional_2021}.
The linearized form captures only the very early time growth of the domains, and the nonlinear terms are required to observe the $t^{1/3}$ passive scaling.
Nonetheless, even in the linearized form, this calculation demonstrates that the leading-order effect of flows appears at $\mathcal{O}(t^{5/4})$, and any amount of surface flows will accelerate the growth scaling beyond the passive scaling. 
While not exact solutions, these trends are consistent with our experimental observations that surface flows can significantly accelerate the kinetics of coarsening.

\begin{acknowledgments}
This material is based upon work supported by the National Science Foundation under Grant No.~2150686.
D.P.A. is supported by the National Science Foundation Graduate Research Fellowship under Grant No.~2139319.
S.C.T. is supported by the Packard Fellowship in Science and Engineering.

\end{acknowledgments}

\end{document}